\documentclass[aps,twocolumn,showpacs]{revtex4}
\usepackage{graphicx}

\begin{document}
\title{Instabilities in numerical loop quantum cosmology}

\author{Jessica Rosen, Jae-Hun Jung and}
\affiliation{Mathematics Department, University of Massachusetts at Dartmouth, North Dartmouth, Massachusetts 02747}

\email{JJung@UMassD.Edu}

\author{Gaurav Khanna}
\affiliation{Physics Department, University of Massachusetts at Dartmouth, North Dartmouth, Massachusetts 02747}

\email{GKhanna@UMassD.Edu}

\date{\today}
\begin{abstract}
In this article we perform von Neumann analysis of the difference equations that arise as a result of loop quantum gravity being applied to models of cosmology and black holes. In particular, we study the numerical stability of Bianchi I LRS (symmetric and non-symmetric constraint) and Schwarzschild interior (symmetric constraint) models, and find that there exist domains over which there are instabilities, generically. We also present explicit evolutions of wave-packets in these models and clearly demonstrate the presence of these instabilities.  
\end{abstract}


\maketitle

\section{Introduction}
Loop quantum cosmology (LQC)~\cite{lqc} is a symmetry-reduced application of loop quantum gravity~\cite{rov}, a theory that leads to space-time that is discrete at Planck scales. An important result of LQC is that it is free of singularities. This is seen as a general consequence of the quantum evolution equation, which is a ``difference equation'' for the wave function and does not break down where the classical singularity would be~\cite{boj02}. In addition, LQC has led to the emergence of a possible mechanism for inflation~\cite{inf}, development of effective semi-classical Hamiltonians~\cite{ban-dat05} and several other interesting directions for research. 

Solving these difference equations that arise in LQC models is an active area of research today. Notions of semi-classicality and the related concept of pre-classicality play a very significant role currently, because in the appropriate limits, solutions to these LQC models must yield the known and expected classical solutions. Recently there has been a lot of progress on this front and several techniques of both analytical~\cite{car-kha} and numerical~\cite{singh}\cite{basis} nature have been developed to obtain appropriate solutions to LQC models. In this article, it is not our focus to further develop methods for obtaining solutions to LQC models. Instead, we investigate a specific property, i.e. the ``generic stability'' of these solutions in certain homogeneous LQC models. Note that a study of a similar type, but in a more general context and emphasis is available in literature~\cite{date}. In this article, we concentrate on Bianchi I locally rotationally symmetric (LRS) cosmology and the Schwarzschild interior geometry.   

While difference equations can be solved very naturally by numerical simulations on computers, it is essential to note one crucial difference between LQC models and the usual finite-difference approaches towards solving and analyzing common differential equations. An important aspect of the numerical analysis of differential equations is the flexibility of  choosing various finite-difference schemes, that in the continuum limit yield the differential equation under consideration. In fact, it is this freedom that is used to develop sophisticated algorithms for numerical evolutions that are stable and convergent. Unfortunately, in numerical LQC we lack this freedom. This is because it is the difference equation itself that is fundamental, and is not the result of a discretization of some differential equation. However, there is some limited flexibility, that is related to quantum ambiguities and different variations of the Hamiltonian constraint in the theory. In this article, we investigate two different forms of the Hamiltonian constraint for Bianchi I LRS model and the dependence on a quantum ambiguity in the Schwarzschild interior geometry model. 

This article is organized as follows. We first present the von Neumann stability analysis of the non-symmetric constraint in the context of the Bianchi I LRS LQC model. Then, we study the same for the Schwarzschild interior LQC model using the symmetric Hamiltonian constraint. We also make some remarks about the Bianchi I LRS symmetric constraint, which is a special case of the Schwarzschild model. Finally, we end with a discussion of our results. 

\section{Numerical stability analysis of LQC models}

\subsection{Bianchi I LRS non-symmetric constraint}

The Bianchi I LRS homogeneous model has two degrees of freedom which are given by two connection components $(A,c)$ and the conjugate momenta $(p_A,p_c)$ with symplectic structure given by $\{A,p_A\}=\frac{1}{2}\gamma\kappa$ and $\{c,p_c\}=\gamma\kappa$. Here, $\kappa=8\pi G$ is the gravitational constant and $\gamma$ the Barbero--Immirzi parameter of loop quantum gravity. The momenta $(p_A,p_c)$ are components of a densitized triad which determine the scale factors $a_I$ of a Bianchi I metric by $a_1=\sqrt{|p_c|}=a_2$, $a_3=p_A/\sqrt{|p_c|}$. Thus, the metric is $d s^2 = |p_c| (dx^2+dy^2)+p_A^2/|p_c| dz^2$ in Cartesian coordinates. 

After quantizing the model with techniques from loop quantum gravity~\cite{homo}, the triad components $p_A$ and $p_c$ become basic operators with discrete spectra, $\hat{p}_A|m,n\rangle=\frac{1}{4}\gamma\ell_P^2 m|m,n\rangle$ and $\hat{p}_c|m,n\rangle=\frac{1}{2}\gamma\ell_P^2 n|m,n\rangle$. The wave function is thus supported on a discrete minsuperspace, and is a solution to the quantized Hamiltonian constraint (non-symmetric) subject to a two-parameter difference equation,

\begin{equation}
 {{u_m^{n+1}-u_m^{n-1}}} = -{{d(n)}\over{2d_2(m)}}
        ({u_{m+1}^n - u_{m-1}^n}),
\label{baby}
\end{equation}
 where 
\begin{eqnarray}
 d_2(m) &=& {1\over {m}}, \quad \mbox{and}\quad d_2(0) = 0,\nonumber \\
 d(n) &=& \sqrt{\left|1+{1\over{2n}}\right|}-\sqrt{\left|1-{1\over{2n}}\right|}, \quad \mbox{and} \quad d(0) = 0. \nonumber
\end{eqnarray}

We shall interpret the discrete parameter, $n$ as a quantum analog of physical ``time'' and the parameter, $m$ as ``space''. We shall focus on the numerical stability of the given difference equation (\ref{baby}). We employ von Neumann analysis which is used to investigate the stability of generic finite difference schemes. The procedure is to perform a (discrete) Fourier transform along all spatial dimensions, thus reducing the system to a recursion relation in time. Using the relation, one computes the ``amplification'' or ``gain'' by calculating the ratio of the values of the transform at consecutive time-steps. If this amplification ratio's absolute value is bounded by unity, the system is stable~\cite{stability}. We begin by writing the above equation in the following fashion. Let,
$$
   v^p_m = u^{2p}_m, \quad \quad w^p_m = u^{2p+1}_m, \quad \mbox{for } \forall m \quad \mbox{and}
   \quad p = 0, 1, \cdots, 
$$
then the equation can be expressed in the form,
$$
 {v_m^{p}-v_m^{p-1}} = -{{d(2p-1)}\over{d_2(m)}}\left(
        {w_{m+1}^{p-1} - w_{m-1}^{p-1}}\right), %\nonumber
$$
and 
$$
 {w_m^{p}-w_m^{p-1}} = -{{d(2p)}\over{d_2(m)}}\left(
        {v_{m+1}^p - v_{m-1}^p}\right), %\nonumber
$$
Now, as prescribed for standard von Neumann stability analysis, we assume that $v^p_m = {\hat v}^p\exp(im\omega)$ and $w^p_m={\hat w}^p\exp(im\omega)$, and then the  difference equations for ${\hat v}^p_m$ and ${\hat w}^p_m$ are given by
$$
     {\hat v}^p -{\hat v}^{p-1} = -{{d(2p-1)}\over{d_2(m)}} 2i\sin\omega{\hat w}^{p-1},
$$
and
$$
     {\hat w}^p - {\hat w}^{p-1} = -{{d(2p)}\over{d_2(m)}}2i\sin\omega{\hat v}^p.
$$
where we used $e^{i\omega} - e^{-i\omega} = 2i\sin\omega$. These can be written in matrix form as seen below,
$$
\left( \begin{array}{cc}
    1 & 0 \\
     {{d(2p)}\over{d_2(m)}}2i\sin\omega&
    1\end{array}\right)
\left(
\begin{array}{cc}
   {\hat v}^p \\ {\hat w}^p \end{array} \right)
 = $$
$$
\left( \begin{array}{cc}
    1 & -{{d(2p-1)}\over{d_2(m)}}2i\sin\omega \\
    0&
    1\end{array}\right)
\left(
\begin{array}{cc}
{\hat v}^{p-1} \\
{\hat w}^{p-1}
\end{array}
\right).
$$
Multiplying each side by 
$
\left( \begin{array}{cc}
    1 & 0 \\
     -{{d(2p)}\over{d_2(m)}}2i\sin\omega&
    1\end{array}\right),
$
the inverse matrix of the LHS matrix yields, 
$$
\left(
\begin{array}{cc}
   {\hat v}^p \\ {\hat w}^p \end{array} \right)
 = $$
$$ 
\left( \begin{array}{cc}
    1 & -{{d(2p-1)}\over{d_2(m)}}2i\sin\omega \\
    -{{d(2p)}\over{d_2(m)}}2i\sin\omega &
    -{{d(2p)d(2p-1)}\over{d_2^2(m)}}4\sin^2\omega +1\end{array}\right)
\left(
\begin{array}{cc}
{\hat v}^{p-1} \\
{\hat w}^{p-1}
\label{babyMatrix}
\end{array}
\right).
$$
Let the coefficient matrix in the RHS of the equation above be ${\bf M}(\omega,p,\mu)$. Then $({\hat v}^p, {\hat w}^p)^T$ becomes
$$
 \left(
\begin{array}{cc}
   {\hat v}^p \\ {\hat w}^p \end{array} \right)
 =\prod_{j=1}^p {\bf M}(\omega,j,m) 
 \left(
\begin{array}{cc}
   {\hat v}^0 \\ {\hat w}^0 \end{array} \right)   
$$
The necessary condition for stability is that absolute values of the eigenvalues of this matrix are bounded by unity. These eigenvalues $\lambda_{1,2}$ of $\bf M$ are
given by
$$ 1 - {1\over 2}D(p,m)\sin^2\omega 
        \pm {1\over 2}\sqrt{\sin^2\omega D(p,m)\left[ D(p,m)\sin^2\omega-4\right]},
$$
where
$$
   D(n,m) = {{d(2p)d(2p-1)}\over{d_2^2(m)}}. 
$$

Now, consider the case, $D(n,m)\sin^2\omega - 4 \le 0$. In this case, is it clear that the absolute value of both eigenvalues presented above, is exactly unity. However, if we consider the complementary case, i.e. $D(n,m)\sin^2\omega - 4>0$ then there do exist eigenvalues with absolute values that are larger than unity, which indicates the presence of an instability in this discrete system. It is interesting to note that it is the larger values of $m$ that induce the instability. In particular, the instability arises if $m$ is in the interval given by (we have converted back to the index parameter, $n$ here)
\begin{equation}
  m^2 \ge {4\over {\sin^2\omega d(n)d(n-1)} },
\end{equation}
where $\sin\omega\ne0$. If $\sin\omega=0$, we have immediately $\lambda = 1$. To see the domain of the instability more clearly, in figure 1 we plot the unstable eigenvalue as a function of $n$ and $m$. Here we simply chose $\sin\omega=1$. The domain of instability is roughly characterized by $m > 4 n$ as seen clearly in the figure and also from the inequality that appears above. 
%%%%%%%%%%%%%%%%%%%%%%%%%%%%%%%%%%%%%%%%%%%%%%%%%%%%%%%%%%%%%%%%%%%%%%%%%%%%
\begin{figure}[ht]
\center
\includegraphics[width=0.4\textwidth]{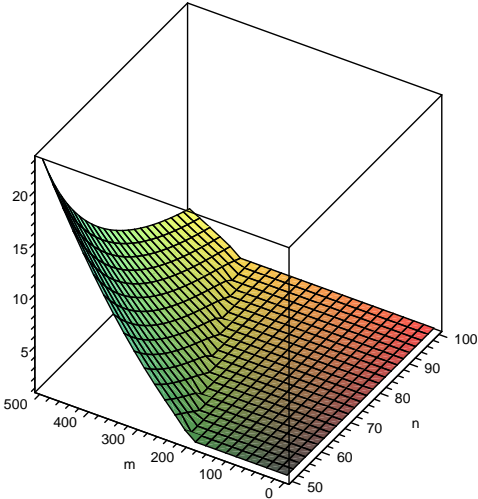}
\caption{\it
Unstable eigenvalue of the ``evolution matrix'' for the Bianchi I LRS LQC model with the non-symmetric Hamiltonian constraint. Notice the range where the eigenvalues
are greater than the unity. 
}
\label{fig1}
\end{figure}

\subsection{Schwarzschild interior model with symmetric Hamiltonian constraint}

Recently, the interior of Schwarzschild geometry was quantized using techniques from LQC~\cite{schwarz}. This is possible because the interior space-time of a Schwarzschild black hole can be considered as a homogeneous cosmology model of Kantowski-Sachs type. The symmetric Hamiltonian constraint in this case, takes the form (for $\tau, \mu > 1$)
\begin{eqnarray}
 &&d_2(\tau)\left(\Psi_{\mu+1}^{\tau+1}-\Psi_{\mu-1}^{\tau+1} \right)-
 d_0(\tau)\left(\Psi_{\mu+1}^{\tau-1}-\Psi_{\mu-1}^{\tau-1} \right) \nonumber \\
 +&&d_1(\tau)\left(m_2(\mu)\Psi_{\mu+2}^{\tau} - \kappa m_1(\mu)\Psi_\mu^\tau +
 m_0(\mu)\Psi_{\mu-2}^\tau\right) = 0 \nonumber
\end{eqnarray}
where

$$
 d_2(\tau) = \sqrt{|\tau+1|}+\sqrt{|\tau|} \nonumber
$$

$$
 d_0(\tau) = \sqrt{|\tau-1|}+\sqrt{|\tau|} \nonumber
$$

$$
 d_1(\tau) = \sqrt{\left|\tau+{1\over2}\right|}-\sqrt{\left|\tau-{1\over2}\right|} \nonumber
$$

$$
 m_2(\mu)= \mu + 1 \nonumber
$$

$$
 m_0(\mu)= \mu - 1 \nonumber
$$

$$
 m_1(\mu)= \mu \nonumber
$$
and $\kappa$ is a parameter that is related to the Immirzi constant and a quantum ambiguity~\cite{schwarz}. In this model, we shall interpret the parameter $\tau$ as ``quantum time'' and $\mu$ as ``quantum space''. Now, let us proceed with von Neumann stability analysis of this difference equation. Let 
$$
\Psi^\tau_\mu = {\hat \Psi}^\tau \exp(i\omega \mu). 
$$
Then the constraint becomes
\begin{eqnarray}
 &&2i\sin(\omega) d_2(\tau){\hat \Psi}^{\tau+1}-
 2i\sin(\omega) d_0(\tau){\hat \Psi}^{\tau-1}\nonumber \\
 +&&d_1(\tau)\left(\mu(2\cos(2\omega)-\kappa)+2i\sin(2\omega)) \right){\hat \Psi}^\tau = 0 \nonumber
 \end{eqnarray}
As before, we define the auxiliary variables $E$ and $B$ such as
$$
     E^t := {\hat \Psi}^t, \quad t = \cdots, -1, 1, 3, \cdots,
$$
and
$$
     B^t := {\hat \Psi}^t, \quad t = \cdots, 0, 2, 4, \cdots. 
$$
We also define the coefficient variables $G_2(\omega,\tau)$, $G_0(\omega,\tau)$ and $H(\omega,\mu)$ such as
$$
     G_2(\omega,\tau) :=  2i\sin(\omega) d_2(\tau),
$$
$$
     G_0(\omega,\tau) := 2i\sin(\omega) d_0(\tau),
$$
and
$$
     H(\omega,\mu) := d_1(\tau)(\mu(2\cos(2\omega)-\kappa)+2i\sin(2\omega)).
$$
Then the difference equation can be rewritten in a matrix representation
$$
\left( \begin{array}{cc}
    G_2(\omega,\tau) & H(\omega,\mu) \\
    0 & 
    G_2(\omega,\tau-1)\end{array}\right)
\left(
\begin{array}{cc}
   E^{\tau+1} \\ B^{\tau} \end{array} \right)
 = $$
$$
\left( \begin{array}{cc}
    G_0(\omega,\tau) & 0 \\
    -H(\omega,\mu) & 
    G_0(\omega,\tau-1)\end{array}\right)
\left(
\begin{array}{cc}
E^{\tau-1} \\
B^{\tau-2}
\end{array}
\right) 
$$
Let the coefficient matrix $M(\omega,\tau,\mu)$ be
$$
   \left( \begin{array}{cc}
    G_2(\omega,\tau) & H(\omega,\mu) \\
    0 & 
    G_2(\omega,\tau-1)\end{array}\right)^{-1}
\left( \begin{array}{cc}
    G_0(\omega,\tau) & 0 \\
    -H(\omega,\mu) & 
    G_0(\omega,\tau-1)\end{array}\right).
$$
Then we have
$$
\left(\begin{array}{cc}
 E^{\tau+1} \\ B^{\tau} \end{array} \right)
 =\prod_{j=1}^\tau M(\omega,j,\mu) 
\left(\begin{array}{cc}
   E^{1} \\ B^{0} \end{array} \right)
$$

The eigenvalue expressions of this matrix are too complicated to present in this article, and will not be especially useful. However, again much like the Bianchi I LRS non-symmetric constraint we discussed before, we find that there are regimes over which the absolute values of the eigenvalues exceed unity and thus indicate the presence of an instability. This happens generically, for any value of $\kappa$. We present below, in figures 2 and 3, plots of the unstable eigenvalue as a function of $\tau$ and $\mu$ and for different values of $\kappa$. Note that the $\kappa=1$ case corresponds to the Bianchi I LRS LQC model with the symmetric constraint. The domain of instability is again roughly characterized by $\mu > 4 \tau$ as seen clearly in the figure. The plots look very similar for other values of $\kappa$, so to save space, we do not include more of those but point out that with larger values of $\kappa$ the unstable region gets progressively larger and the instability gets worse (see figure 4). For the sake of completeness, we also show a plot of the stable eigenvalue for the case of $\kappa=2$ in figure 5.    
%%%%%%%%%%%%%%%%%%%%%%%%%%%%%%%%%%%%%%%%%%%%%%%%%%%%%%%%%%%%%%%%%%%%%%%%%%%%
\begin{figure}[ht]
\center
\includegraphics[width=0.4\textwidth]{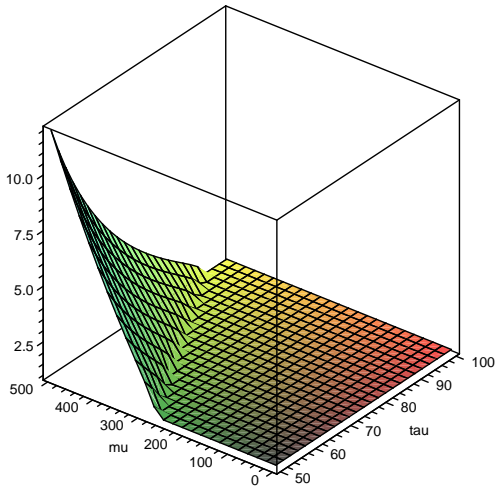}
\caption{\it
Unstable eigenvalue of the ``evolution matrix'' for the Bianchi I LRS LQC model with the symmetric Hamiltonian constraint. Notice the range where the eigenvalues
are greater than the unity. 
}
\label{fig2}
\end{figure}
%%%%%%%%%%%%%%%%%%%%%%%%%%%%%%%%%%%%%%%%%%%%%%%%%%%%%%%%%%%%%%%%%%%%%%%%%%%%
\begin{figure}[ht]
\center
\includegraphics[width=0.4\textwidth]{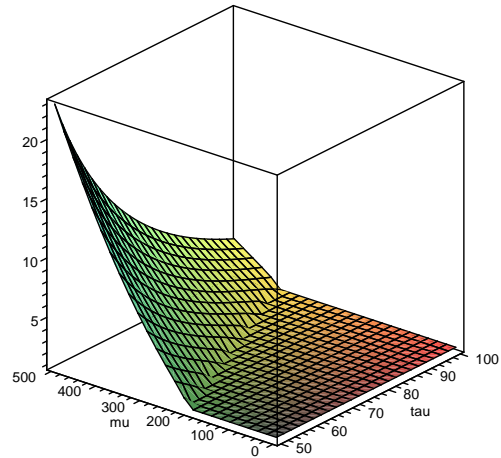}
\caption{\it
Unstable eigenvalue of the ``evolution matrix'' for the Schwarzschild interior LQC model with the symmetric Hamiltonian constraint ($\kappa=2$). Notice the range where the eigenvalues are greater than the unity. 
}
\label{fig3}
\end{figure}
%%%%%%%%%%%%%%%%%%%%%%%%%%%%%%%%%%%%%%%%%%%%%%%%%%%%%%%%%%%%%%%%%%%%%%%%%%%%
\begin{figure}[ht]
\center
\includegraphics[width=0.4\textwidth]{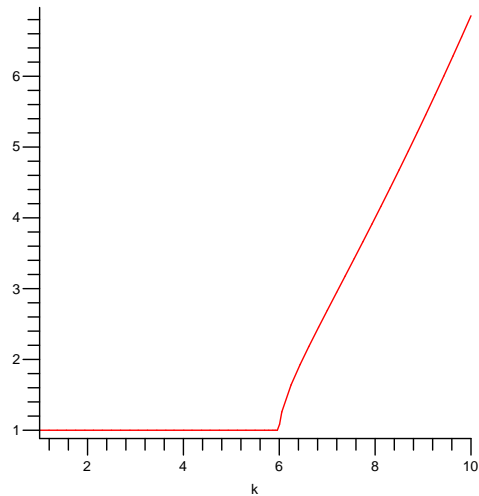}
\caption{\it
Eigenvalue plotted as a function of $\kappa$ at fixed values of $\mu=\tau=100$. Clearly, larger values of $\kappa$ make the system more unstable and also make the unstable region become progressively larger.  }
\label{fig4}
\end{figure}
%%%%%%%%%%%%%%%%%%%%%%%%%%%%%%%%%%%%%%%%%%%%%%%%%%%%%%%%%%%%%%%%%%%%%%%%%%%%
\begin{figure}[ht]
\center
\includegraphics[width=0.4\textwidth]{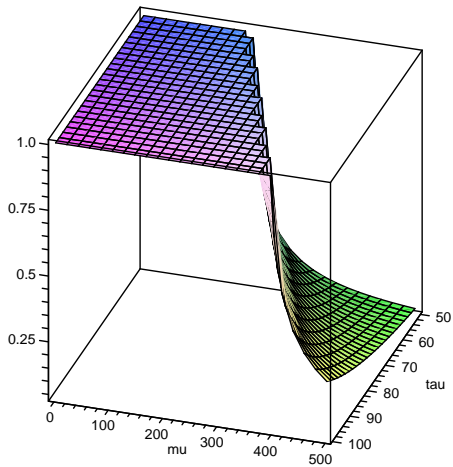}
\caption{\it
Stable eigenvalue of the ``evolution matrix'' for the Schwarzschild interior LQC model with the symmetric Hamiltonian constraint ($\kappa=2$). }
\label{fig5}
\end{figure}

\subsection{Evolution of wave-packets}

Below in figure 7 and 8, we also present explicit numerical evolutions of wave-packets for the $\kappa=2$ case. These evolutions were performed on the basis of the stencil depicted in figure 6, which shows the values of the wave function that are related by the Hamiltonian constraint in consideration here. Also note that there are some residual gauge symmetries that indicate that the wave function is unaffected by the gauge transformation $\mu \to -\mu$, coming from the Gauss constraint~\cite{schwarz}. Putting this together, we can fix the boundaries of our numerical simulation to be $\mu = \pm M$, for large M. Further, there are arguments~\cite{schwarz} for choosing $\Psi \to 0$ as $\mu \to \infty$. Thus, as long as the maximum grid size $M$ is large enough to avoid sizeable errors in solving the difference equation, we can set $\Psi_{\pm M}^\tau = 0$. 

Our results are as expected. If the wave-packet is evolved over a domain where the evolution is expected to be unstable, we clearly see that the numerical evolution fails after a few time-steps. On the other hand, over the stable domain we appear to be able to numerically evolve the wave-packets for a long time with success. 
%%%%%%%%%%%%%%%%%%%%%%%%%%%%%%%%%%%%%%%%%%%%%%%%%%%%%%%%%%%%%%%%%%%%%%%%%%%%
\begin{figure}[ht]
\center
\includegraphics[width=0.4\textwidth]{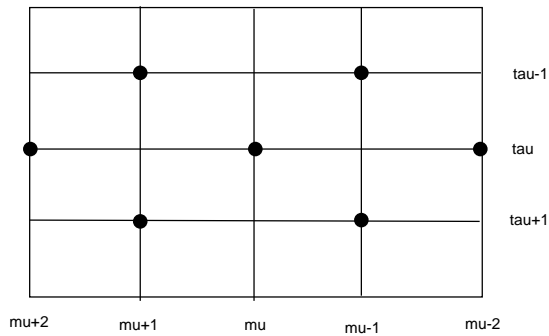}
\caption{\it
Stencil, that shows the values of the wave function that are related by the Schwarzschild interior or Bianchi I LRS symmetric Hamiltonian constraint. }
\label{fig6}
\end{figure}
%%%%%%%%%%%%%%%%%%%%%%%%%%%%%%%%%%%%%%%%%%%%%%%%%%%%%%%%%%%%%%%%%%%%%%%%%%%%
\begin{figure}[ht]
\center
\includegraphics[width=0.4\textwidth]{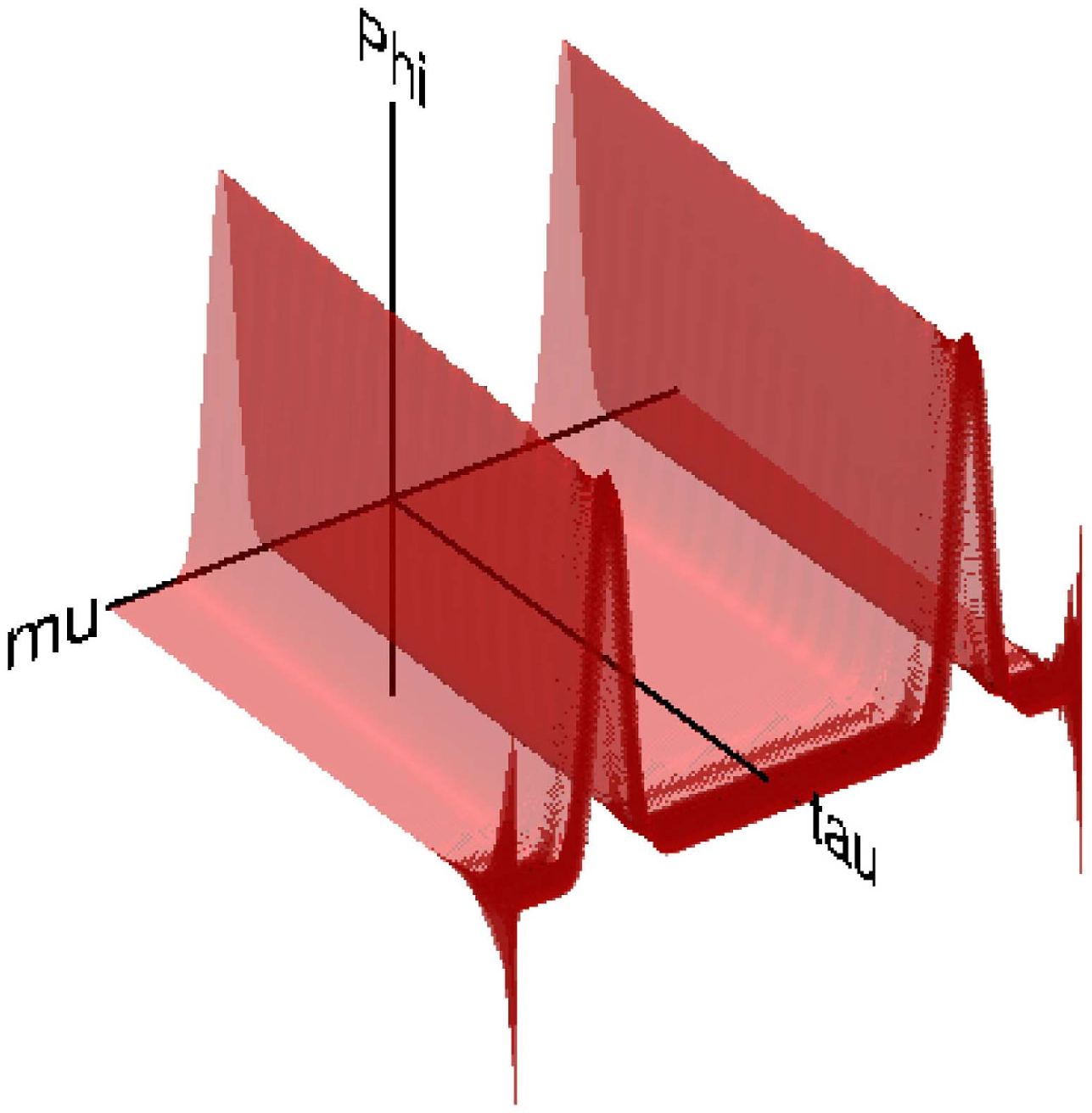}
\caption{\it
Unstable wave-packet numerical evolution for the Schwarzschild interior LQC model with the symmetric Hamiltonian constraint ($\kappa=2$). We chose the range of $\mu$ to $(-1000,1000)$ and we evolved forward in $\tau$ from $150$ to $186$. The initial wave-packets were centered at $\pm 600$ and had a width of $50$ which places part of them in the unstable region.}
\label{fig7}
\end{figure}
%%%%%%%%%%%%%%%%%%%%%%%%%%%%%%%%%%%%%%%%%%%%%%%%%%%%%%%%%%%%%%%%%%%%%%%%%%%%
\begin{figure}[ht]
\center
\includegraphics[width=0.4\textwidth]{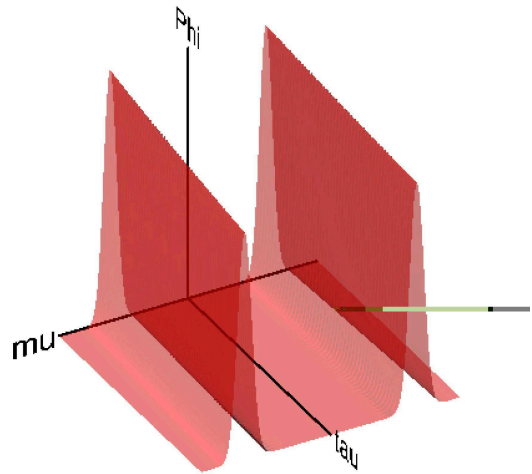}
\caption{\it
Stable wave-packet evolution for the Schwarzschild interior LQC model with the symmetric Hamiltonian constraint ($\kappa=2$). We chose the range of $\mu$ to $(-1000,1000)$ and we evolved forward in $\tau$ from $200$ to $300$. The initial wave-packets were centered at $\pm 600$ and had a width of $50$ which places them in the stable region.}
\label{fig8}
\end{figure}

\section{Discussion}

In this article, we performed von Neumann numerical stability analysis of some of the difference equations that arise in homogeneous LQC models. In particular, we focussed on Bianchi I LRS cosmology and the Schwarzschild interior geometry. We investigated two different forms of the Hamiltonian constraint for Bianchi I LRS model and the dependence on a quantum ambiguity in the Schwarzschild interior geometry model. In all these cases, we discovered that there are large regions in space-time that have instabilities, generically. These instabilities are characterized by rapid exponential growth, and therefore they are difficult to interpret physically. The presence of these instabilities, especially in the large volume limit is troubling because it is indicative of the fact that the models do not have the correct semi-classical behavior. 

These results may be related to another recent troubling result from the study of pre-classical solutions of the Schwarzschild interior LQC model using generating function techniques~\cite{bound}. There, an upper bound was discovered on the quantity we called $\kappa$ here, which in turn led to a bound on the value of the Immirzi parameter. This bound on the Immirzi parameter appears to be violated by its currently accepted values. It is possible that our current results are simply further indicating that these models appear to be lacking in some fashion. 

Another possibility is that these regions that appear to have the instability are ``forbidden'' regions in some manner. What is interesting about this possibility is that these regions are forbidden quantum mechanically, but not forbidden classically (see classical trajectories of this LQC model in~\cite{schwarz}). If this would be true, it would be a very unusual scenario. We do not feel that this is a very likely possibility. 

Finally, we'd like to comment on yet another interesting possibility~\cite{new}. Note that a region is semi-classical if the length and time scales are large and the curvature scales are small. In mini-superspace models, such as the ones under consideration here, the relevant curvature can become large with the spatial size. This creates the kind of instability problems we're observing here. Moreover, difference equations that are derived in the context of mini-superspace models do not take into account the subdivision of discrete geometry which is expected when the Universe grows larger. Taking this into account requires introducing inhomogeneity or introducing a scale-dependent discretization, which is currently a topic of exploration. However, even in these cases its is possible that unstable regions may not go away completely and thus may still present interesting opportunities for further research.    

\section{Acknowledgments}

We appreciate comments and suggestions from Daniel Cartin, Martin Bojowald and Jorge Pullin. We acknowledge research support from the Mathematics and Physics departments of the University of Massachusetts at Dartmouth. GK acknowledges support from Glaser Trust of New York.

\end{document}